\documentclass{iaus}
\usepackage{graphicx}

\title[Galactic Planetary Nebulae] 
{A Consolidated Online Database\\ of Galactic Planetary Nebulae}

\author[B. Miszalski, A. Acker, F. Ochsenbein \& Q.A. Parker]   
{Brent Miszalski$^{1,2}$, A. Acker$^{3}$, F. Ochsenbein$^{3}$ \& Q. A. Parker$^{4,5}$
}

\affiliation{
$^1$South African Astronomical Observatory\ 
$^2$Southern African Large Telescope Foundation
\\email: {\tt brent@saao.ac.za}\\[\affilskip]
$^{3}$Observatoire astronomique de Strasbourg\\email: {\tt agnes.acker@astro.unistra.fr}\\[\affilskip]
$^{4}$Macquarie University\ 
$^{5}$Australian Astronomical Observatory
}

\pubyear{2011}
\volume{283}  
\pagerange{100-104}
\jname{Planetary Nebulae an Eye to the Future}
\editors{A.C. Editor, B.D. Editor \& C.E. Editor, eds.}
\begin{document}

\maketitle

\begin{abstract}
   Since the unifying Strasbourg-ESO Catalogue of Galactic Planetary Nebulae\\ (SECGPN) a large number of new discoveries have been made thanks to improved surveys and discovery techniques. The increasingly heterogeneous published population of Galactic PNe, that we have determined totals $<$ 2850 PNe, is becoming more difficult to study on the whole without a centralised repository. We introduce a consolidated and interactive online database with object classifications that reflect the latest multi-wavelength data and the most recent results. The extensible database, hosted by the Centre de Donnees astronomique de Strasbourg (CDS), will contain a wealth of observed data for large, well-defined samples of PNe including coordinates, multi-wavelength images, spectroscopy, line intensities, radial velocities and central star information. It is anticipated that the database will be publicly released early 2012.
\end{abstract}

\firstsection 
              
\section{How many Galactic PNe are there?}
\begin{figure}
   \begin{center}
      \includegraphics[scale=0.6,angle=270]{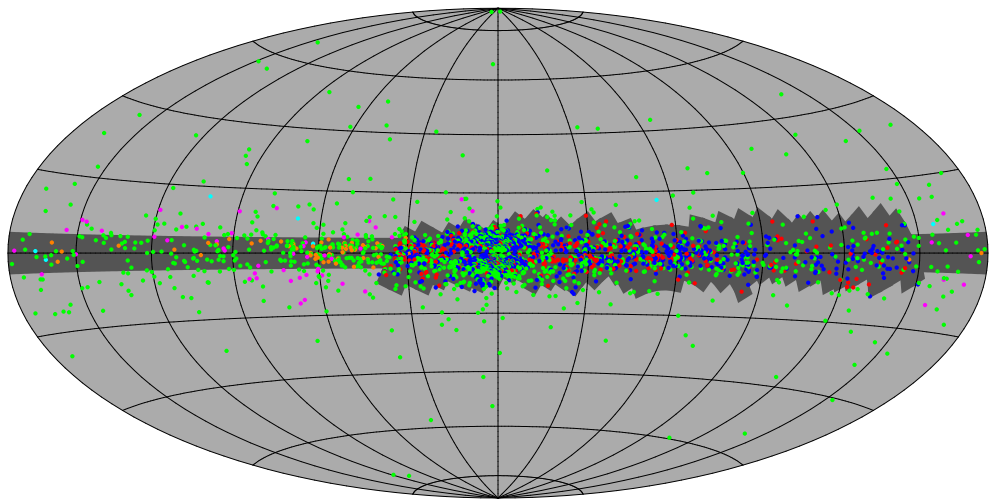}
   \end{center}
   \caption{Current database contents in a Galactic Hammer projection. Major published catalogues are coloured separately to most PNe (green) and include MASH (blue, Parker et al. 2006), MASH-II (red, Miszalski et al. 2008), IPHAS (orange, Viironen et al. 2009), Deep Sky Hunters (magenta, Jacoby et al. 2010) and the first few discoveries from ETHOS (cyan, Miszalski et al. in prep). The darker grey bands represent the IPHAS (Drew et al. 2005) and SHS (Parker et al. 2005) survey footprints.}
   \label{fig:fig1}
\end{figure}

The database was meticulously constructed by visually inspecting images of purported PNe (sourced from major catalogues, SIMBAD and other literature sources) in the SuperCOSMOS Halpha Survey (SHS, Parker et al. 2005), the INT Photometric Halpha Survey (IPHAS, Drew et al. 2005) and the SuperCOSMOS Sky Surveys (SSS, Hambly et al. 2001). During this process both PNe and obvious non-PNe (e.g. Galaxies, photographic flaws, HII regions, duplicates, etc.) were added to the database. The non-PNe were flagged as different object types to keep them separate and to keep track of objects already considered for inclusion in the database. This laborious process allowed us to make the most accurate estimate to date of the total number of Galactic PNe, $N_\mathrm{gal}$, instead of the usual approach of adding total numbers found in new catalogues. We find $N_\mathrm{gal} < 2850$, though this upper limit may be reduced as more non-PNe are identified after we further check and refine our database before its first release. Our value is based on currently published catalogues and will also be revised upwards once future PN discoveries are published. Figure \ref{fig:fig1} depicts the current Galactic distribution of PNe.

\firstsection 
\section{Database interface}
The database will be accessible through a web-based interface that allows users to search for PNe by object name (IAU PN G and common name) and J2000 coordinates (single or list format). Static releases will also be made intermittently to VizieR for virtual observatory compatibility. Results from the web-interface may be browsed in a gallery format with thumbnail images of each PN, as well as a variety of other formats. Groups of PNe may be selected based on their coordinates, object size, name, catalogue and so on. Links to online resources are also made available including e.g. the ESO archive, VizieR and Staralt.

\section{The First Release}
The first release of the database is planned for early 2012 and is intended to produce a working database with the cleanest and largest set of entries for published PNe. 
This will allow for queries to be made, selecting large samples of PNe to be studied in a unified fashion with accurate coordinates and PN G designations. 
The initial dataset will include at least image cut-outs from the SHS and SSS (in \textsc{jpeg} and \textsc{fits} format), SECGPN spectra in \textsc{fits} format (Acker et al. 1992), radial velocities (e.g. Durand et al. 1998) and central star magnitudes (e.g. Tylenda et al. 1991). Additions to the database will mostly include large observational datasets in refereed publications that are of broad interest to the whole sample.


\begin{thebibliography}{}
\bibitem[Acker et al.(1992)]{1992secg.book.....A} Acker, A., Marcout, J., Ochsenbein, F., Stenholm, B., \& Tylenda, R.\ 1992, Garching: European Southern Observatory, 1992
\bibitem[Drew et al.(2005)]{2005MNRAS.362..753D} Drew, J.~E., Greimel, R., Irwin, M.~J., et al.\ 2005, MNRAS, 362, 753 
\bibitem[Durand et al.(1998)]{1998A&AS..132...13D} Durand, S., Acker, A., \& Zijlstra, A.\ 1998, A\&AS, 132, 13 
\bibitem[Hambly et al.(2001)]{2001MNRAS.326.1279H} Hambly, N.~C., MacGillivray, H.~T., Read, M.~A., et al.\ 2001, MNRAS, 326, 1279 
\bibitem[Jacoby et al.(2010)]{2010PASA...27..156J} Jacoby, G.~H., Kronberger, M., Patchick, D., et al.\ 2010, PASA, 27, 156 
\bibitem[Miszalski et al.(2008)]{2008MNRAS.384..525M} Miszalski, B., et al.\ 2008, MNRAS, 384, 525 
\bibitem[Parker et al.(2005)]{2005MNRAS.362..689P} Parker, Q.~A., Phillipps, S., Pierce, M.~J., et al.\ 2005, MNRAS, 362, 689 
\bibitem[Parker et al.(2006)]{2006MNRAS.373...79P} Parker, Q.~A., et al.\ 2006, MNRAS, 373, 79 
\bibitem[Tylenda et al.(1991)]{1991A&AS...89...77T} Tylenda, R., Acker, A., Raytchev, B., Stenholm, B., \& Gleizes, F.\ 1991, A\&AS, 89, 77 
\bibitem[Viironen et al.(2009)]{2009A&A...504..291V} Viironen, K., et al.\ 2009, A\&A, 504, 291 

\end{thebibliography}
\end{document}